\input harvmac
\pretolerance=10000

\Title{HWS-9821,hep-th/0001008}
{\vbox{\centerline{$N=0$ Supersymmetry and the Non-Relativistic
Monopole}}}

\centerline{Donald Spector\footnote{$^\dagger$}{spector@hws.edu} }
\medskip\centerline{Department of Physics, Eaton Hall}
\centerline{Hobart and William Smith Colleges}
\centerline{Geneva, NY \ 14456 USA}

\vskip .3in
We study some of the algebraic properties of the non-relativistic
monopole.  We find that we can
construct theories that possess an exotic
conserved fermionic charge that squares to the Casimir of the  rotation
group,
yet do not possess an ordinary supersymmetry.
This is in contrast to previous known examples with such exotic
fermionic charges.  
We proceed to show that the presence of the exotic fermionic
charge in the non-supersymmetric
theory can nonetheless
be understood using supersymmetric techniques, providing
yet another example of the usefulness of supersymmetry in understanding
non-supersymmetric theories.

\Date{12/99}

\newsec{Introduction}

The use of supersymmetry to understand non-supersymmetric theories
has proven to be a valuable resource in the study of quantum 
theories\ref\shape{F. Cooper, A. Khare, and U. Sukhatme, {\it Phys.
Rep.} {\bf 251} (1995) 267.}\ref\zhdas{
Z. Hlousek and D. Spector, {\it Mod. Phys. Lett.} {\bf A13} (1998)
202\semi
Z. Hlousek and D. Spector, {\it Nucl. Phys.} {\bf B442} (1995) 413\semi
Z. Hlousek and D. Spector, {\it Nucl. Phys.} {\bf B397} (1993) 173\semi
Z. Hlousek and D. Spector, {\it Mod. Phys. Lett.} {\bf A7} (1992) 3403.
}\ref\scat{
 F.A. Berends, W.T. Giele, and H. Kuijf, {\it Phys. Lett.} {\bf 211B}
(1988) 91\semi
D. Kosower, B.-H. Lee, and V.P. Nair, {\it Phys. Lett.} {\bf 201B}
(1988)
85\semi
S. Parke, M. Mangano, and Z. Xu, {\it Nucl. Phys.} {\bf B298} (1988)
653\semi
S. Parke and T.R. Taylor, {\it Phys. Lett.} {\bf 157B} (1985) 81.
}.  In the
examples just listed, one typically 
finds a way to understand a theory that is 
not supersymmetric by treating it as the restriction of a
supersymmetric theory, typically by eliminating the fermionic
fields of the supersymmetric theory..

There is, of course, another possibility.  Suppose, instead,
the supersymmetric
theory is the restriction of some non-supersymmetric theory.
Can the algebraic structure of the supersymmetric theory nonetheless
give an indication as to the behavior of the non-supersymmetric theory?
In this letter, we study an example where this occurs.

The importance of this is twofold.  First, we obtain some particular
insights into the theory of non-relativistic magnetic monopoles 
and dyons.  But,
more importantly, we  extend the usefulness of supersymmetry in
understanding non-supersymmetric theories.  The analysis of quantum 
systems is sufficiently difficult that any new techniques are useful.
Because supersymmetry is itself so powerful, any time we can link a 
non-supersymmetric theory to a supersymmetric one, we
have the potential of
deeper insights into the non-supersymmetric theories.

In the first section of this
paper, we present a non-supersymmetric model that has
an exotic conserved
fermionic charge.  Existing examples of the appearance of such
charges have always been in supersymmetric 
theories \ref\exoticsusy{F. De Jonghe, A.J. Macfarlane,
K. Peeters, and J.W. van Holten, {\it Phys. Lett.} {\bf 359B}
(1995) 114\semi
A.J. Macfarlane, {\it Nucl. Phys.} {\bf B438} (1995) 455\semi
J.W. van Holten, {\it Phys. Lett.} {\bf 342B} (1995) 47.}, 
and with the 
supersymmetry algebra invoked in an essential way to explain
appearance of such exotic charges \ref\skysusy{G.W. Gibbons,
R.H. Rietdijk, and J.W. van Holten, {\it Nucl. Phys.} {\bf B404}
(1993) 42.}.  
Thus,
in this section, we establish a counterexample, in which an exotic
fermion
charge (one that does not square to the Hamiltonian) can exist in
a context in which there is no ordinary supersymmetry, and so the
charge cannot be understood in the usual way.

Then we examine the algebraic structure of this theory in some
depth.  We find an intriguing set of conjugate operators that lead
us to an interesting formulation of the theory, one that affords
a straightforward way to motivate the identification of
the exotic fermionic charge. We comment that it is also possible
that these operators can be used to demonstrate integrability in
some cases (for a discussion of integrability in non-relativistic
monopole theories, see \ref\intjackiw{R. Jackiw, {\it Ann. Phys.}
{\bf 129}
(1980) 183.}), but that is beyond the scope of this work.

Finally, we use the algebraic properties of these operators to analyze
the appearance of the exotic fermionic charge in the non-supersymmetric
theory.  We find that it is possible to draw a connection between
the non-supersymmetric theory and the supersymmetric theory, and thus
explain the appearance of the exotic charge using supersymmetric 
arguments, even though the theory in question is not supersymmetric!  We
will thus have established an $N=0$ supersymmetry approach to these
exotic charges.

\newsec{A Non-Supersymmetric Monopole Model}

Consider the non-relativistic theory in three spatial dimensions
of a spin ${1\over 2}$ particle in the presence of a magnetic monopole
or dyon.  Such a theory has a Hamiltonian
\eqn\hamone{H =
{1\over 2m}(\vec p-e\vec A)^2 - {1\over 2}e
F^{ij}S^k\epsilon_{ijk}+V(r)~~~,}
where $A^i$ is the gauge field of a magnetic monopole, 
${1\over 2}\epsilon^{ijk}F_{ij}$ is the associated
magnetic field, $S^k$ is the spin operator for the particle,
$e$ is its electric charge,
and $V(r)$ is a
spherically symmetric potential
energy (which will include a Coulomb term if the monopole field
arises from a dyon). The radial coordinate $r=(x_ix_i)^{1\over 2}$. One
typical parametrization of the monopole gauge field in 
spherical coordinates is

\eqn\monoAfieldI{
   \vec A^{(I)}=
   {g\over 4\pi }\,{1-\cos\theta\over r\, \sin\theta}
   \,\hat\phi
   ={g\over 4\pi r}\,\tan{\theta\over2}
   \,\hat\phi \qquad,\qquad \theta<\pi}

 and 
\eqn\monoAfieldII{
   \vec A^{(II)}=
   -{g\over 4\pi}\,{1+\cos\theta\over r\, \sin\theta}
   \,\hat\phi=
   -{g\over 4\pi r}\,\cot{\theta\over2} 
   \,\hat\phi  \qquad,\qquad \theta>0}
where $g$ is the magnetic charge of the configuration.  These
two expressions differ by a gauge transformation in the region of
overlap.

Of course, the angular momentum is conserved in this theory, although
it picks up an anomalous term from the monopole field.  
Defining the covariant momentum operators $\Pi^i = p^i - e A^i$,
we can write the conserved angular momentum operators as
\eqn\angmom{J^i = \epsilon^{ijk}x_j\Pi_k +S^i - eg{x^i\over r}~~~.}

It is useful to re-parametrize the spin in terms of a Grassman 
coordinate $\psi_i$, $i=1,2,3$. 
The $\psi_i$ satisfy
\eqn\psianti{\Bigl\{\psi_i,\psi_j\Bigr\} = \delta_{ij}~~~,}
and they are introduced into the Hamiltonian via the
identification
\eqn\Sandpsi{S^i = -{i\over 2}\epsilon^{ijk}\psi_j\psi_k~~~.}
The $S^i$ satisfy the spin ${1\over 2}$ commutation relations.
These Grassman variables will be central to the analysis throughout
this paper.

Suppose now we define the fermionic charge $\tilde Q$ via the
expression
\eqn\defqtilde{\tilde Q 
= \epsilon_{ijk}\Bigl( x^i\Pi^j\psi^k - 
{i\over 3} \psi^i\psi^j\psi^k\Bigr)~~~.}
It is a straightforward computation to show that
\eqn\consqtilde{\Bigl[H,\tilde Q\Bigr] = 0~~~,}
leading to the conclusion that there is an
exotic conserved fermionic
charge in this theory.

This charge, however, is not an ordinary supercharge.  It does not
square to the Hamiltonian (indeed, dimensionally, it could not).
Its square is, however, a bosonic charge
already known to be present in the theory.
One immediately verifies that
\eqn\qtildesq{\tilde Q^2 =
 {1\over 2}\bigl(J^2  - e^2g^2 +{1\over 4}\bigr).}
This is important, in that it tells us that $\tilde Q$
 does not lead to a whole new
elaborate symmetry structure; rather, this exotic fermionic
charge fits inside the standard symmetry structures
in the minimal possible way.

Such a charge was
observed in \exoticsusy\ in 
the context
of a supersymmetric theory, and
explained as arising from the interplay
of the standard supersymmetry and a Killing-Yano structure
in the theory, a special instance 
of \skysusy.  However, here we have found such a charge
in  the absence of supersymmetry, and so the explanation of
\exoticsusy\ cannot be adequate.  Thus we are left to seek the
origin of this charge in this
non-supersymmetric theory.  How could we have known to look
for it?  Why should it appear?

\newsec{Algebraic Relations among Operators}

Before we attempt to establish a connection between the supersymmetric
and non-supersymmetric theories, we wish to start by exploring some of 
the properties of operators that will turn out to be relevant to both
theories.  From the properties of these operators, we will be able
to motivate the discovery 
and construction of the $\tilde Q$ charge in the non-supersymmetric
theory. 

Suppose there is an exotic fermion charge in a non-relativistic
quantum theory.  What will its properties be?

Let us consider the simplest possibility.  If there is to be such
a charge, the simplest possibility is that there is only one such
charge,
and hence it must be a scalar
under the rotation group.  Its square must then be a conserved
bosonic scalar charge, and since this must not be the Hamiltonian
(we are attempting to construct an exotic fermionic charge, not a
conventional supercharge), if we
are to add no more structure than necessary, the
fermionic charge should square (up to irrelevant 
constants) to the
Casimir of the rotation group.  Then, since
$\bigl[J^2,f(r)\bigr]=0$ for
any function $f(r)$, we also have $\bigl[\tilde Q^2,f(r)\bigr]=0$,
and then the Jacobi identity gives
\eqn\tildeqfr{ \Bigl\{ \tilde Q, \Bigl[ \tilde Q, f(r)\Bigr]\Bigr\} =
0~~~.}
The simplest way for this to be true for any $f(r)$
is for $\tilde Q$ to commute with $r$.

Now there turns out to be a fermionic precursor of $r$ in this
theory.  Let us define $\Gamma = x\cdot\psi$.  Then $\Gamma^2 = r^2$,
and hence we could adopt the ansatz $\bigl\{\tilde Q,\Gamma\bigr\}=0$.
Then \tildeqfr\ is naturally satisfied.

The Hamiltonian framework makes it natural to consider
the conjugate operator to $\Gamma$, namely $Q=\Pi\cdot\psi$.
For convenience, we also define
$W=\{Q,\Gamma\}$; one
sees readily that $W$ essentially measures the engineering dimension of
an operator.

We know that the essential ingredient of quantum mechanics comes
from the action of $x^i$ on $\Pi^j$, or equivalent of $\Pi^j$ on
$x^i$.  What happens
if we consider the corresponding fermi-contracted 
operators here?

If we consider
the repeated action of $Q$ on operators, starting with $\Gamma^2=r^2$, 
one finds
\eqn\Qarrow{Q: \Gamma^2\rightarrow\Gamma\rightarrow 
W\rightarrow Q\rightarrow Q^2\rightarrow 0~~~.} 
On the other hand, we can switch the roles of $\Gamma$ and $Q$,
and then 
\eqn\Gammaarrow{\Gamma: Q^2\rightarrow Q\rightarrow
W\rightarrow\Gamma\rightarrow\Gamma^2\rightarrow 0~~~.}
(In these expressions, we have omitted overall normalization
factors
that are irrelevant to our argument.) 
One notices a complete parity between these two chains.  Under the
action
of $Q$, one starts at $\Gamma^2$ and proceeds all the way to $Q^2$
before
reaching zero; under the action of $\Gamma$, one starts at $Q^2$ and
proceeds all the way to $\Gamma^2$ before reaching zero.  The roles
of $Q$ and $\Gamma$ are switched.  Note, too, that they act in 
reverse directions, each one (roughly speaking) undoing the action
of the other.

Now let us again consider $\tilde Q$.  If it is to square to
the Casimir of the rotation group, it should have engineering
dimension zero, and thus should be composed of terms that have
an equal number of $x^i$'s and $\Pi^i$'s.  Indeed, the charge
$\tilde Q$ should be unchanged under the canonical 
transformation $x^i\rightarrow \Pi^i$, $\Pi^i\rightarrow -x^i$.
Now we have already argued that it is natural to have $\tilde Q$
commute with $\Gamma$.
Given the symmetry between
the algebraic relations \Qarrow\ and \Gammaarrow, it is
natural to conjecture, then, that any conserved charge $\tilde Q$ 
will also commute with $Q$.

Thus, a natural way to try to identify any possible exotic fermionic
charge is to construct a dimension zero operator, unchanged under
the canonical transformation described above, that commutes with
$\Gamma$ and $Q$.  In fact, the charge $\tilde Q$ defined in the
previous section meets all these criteria.  As we will see in the
next section, it is exactly these properties that forge the link between
the supersymmetric and non-supersymmetric theories.

\newsec{Connection to the supersymmetric theory}

We can re-write the Hamiltonian in a more compact, and more instructive,
form.  Using the covariant derivative $\Pi^i$,
one can write the Hamiltonian \hamone\ as
\eqn\hamtwo{H = \Bigl(\Pi\cdot\psi\Bigr)^2 + V(r)~~~.}
Then using the previously defined
operator $Q = \Pi\cdot\psi$, and introducing a parameter
$\alpha$ so we can adjust the magnitude of the potential term,
we are led to the Hamiltonian
\eqn\Halpha{H_\alpha = Q^2 + \alpha V(r)~~~.}
Note that
the case $\alpha = 0$, which describes
a spin-${1\over 2}$ particle moving only in the field
of a magnetic monopole, is thus an example of
a supersymmetric Hamiltonian, with $Q=\Pi\cdot\psi$
as the supercharge
\ref\fardho{E. d'Hoker and L. Vinet, {\it Phys. Lett.}
{\bf 137B} (1984) 72.}.  When $\alpha\ne 0$,
we recover the generic case this paper has been considering,
which has no supersymmetry.

Consider first the theory with Hamiltonian $H_0 = Q^2$, which
arises when one sets $\alpha=0$.
This theory automatically has an exotic fermionic charge, as
demonstrated
in \exoticsusy.  This is automatic, because the theory
has a
Killing-Yano tensor.  From this, the appearance of the
conserved $\tilde Q$ follows automatically by considering the
action of the ordinary supersymmetry on the
Killing-Yano
tensor \skysusy.  This explanation says that the
conserved charge $\tilde Q$ appears in part because
of the existence of the ordinary conserved supercharge.

How do we explain, then, the existence of a conserved $\tilde Q$ when
$\alpha\ne 0$ and there is no ordinary supersymmetry?  How does this
non-supersymmetric extension of the original theory preserve this
one aspect of the symmetry structure?

The answer is that adding the spherically symmetric potential deforms
the theory in just the right way.  It is true that it is a deformation
that violates the supersymmetry invariance of the theory, and thus the
Killing-Yano argument for the appearance of the extra fermionic charge
breaks down.  However, although this deformation violates supersymmetry,
it does commute with $\tilde Q$, and
thus
we are deforming a theory which {\it has} a natural
supersymmetric explanation for
the conservation of $\tilde Q$ in a way that, although
it violates supersymmetry,
respects the $\tilde Q$ conservation law.  
Thus the non-supersymmetric extension 
preserves some of the algebraic structure of the original supersymmetric
theory, in particular, the existence of the
exotic fermionic conserved charge.

Thus arguments based on the supersymmetry algebra can explain the
appearance of $\tilde Q$ as a conserved charge in the non-supersymmetric
theory; we thus have
identified another instance of so-called
``$N=0$ supersymmetry,'' in which supersymmetry is used to determine
the properties of a non-supersymmetric theory.  In fact, in this case,
the term is especially apt. The phenomenon we are witnessing is much
like
that which occurs when new terms are added to an $N=2$ supersymmetric
theory,
breaking one but not both of the supersymmetries, thereby reducing the
invariance to $N=1$ supersymmetry.  Here we are seeing the same sort of
reduction in the number of supersymmetries, although it is a reduction
from
$N=1$ to $N=0$.

\newsec{Conclusion}

We have seen that the appearance of an exotic fermionic conserved
charge --- one that does not square to the Hamiltonian, as would
an ordinary supersymmetry charge --- can occur in a non-supersymmetric
theory.  Heretofore, such exotic charges had been found and explained
in explicitly supersymmetric contexts.

At the same time, we have seen that the appearance of this charge
in the non-supersymmetric theory can still be understood by examining
the algebraic structure of the theory, and in particular by
understanding
how the non-supersymmetric theory can be viewed as a particular
kind of deformation of a supersymmetric theory.

The above is yet another nice example of a way to link 
the behavior of supersymmetric
and non-supersymmetric theories, and indeed it parallels very nicely
the discussion of extended superalgebras and topological charges
in \zhdas.  In both cases, one has a quantity
(the topological charge or $\tilde Q$, respectively) that is conserved
in the supersymmetric and non-supersymmetric cases; this quantity
is conserved in the supersymmetric case
due to the interplay of the supercharge and a geometrical
quantity (the gauge-like potential or the Killing-Yano tensor,
respectively);
and the change to the non-supersymmetric case preserves the
conservation law (due to its being topological or due to the nature of
the
deformation, respectively).  It would be interesting to see if it is
more than
coincidental that both these examples revolve around the supermultiplet
of
geometrical structures associated with symmetries and conservation 
laws that arise in the presence
of monopoles.\foot{There is also a certain similarity 
to \ref\solstat{D. Spector, {\it Mod. Phys. Lett.}
{\bf A9} (1994) 2245.}, in which a supersymmetric theory
is deformed in two different ways, one that respects and one that
violates
supersymmetry, and the properties of these different theories are
related to each other.}

There is another possibility raised by the symmetry between the
chains of operator 
transformations \Qarrow\ and \Gammaarrow, namely the possibility of a
simple
proof of integrability for non-relativistic monopole systems.
The symmetry between these chains even
when the Hamiltonian does not have a symmetry under the interchange
of $\Gamma$ and $Q$ suggests a possible tool for the 
construction of a second Hamiltonian
structure, and thus a proof of integrability. The investigation of
this topic goes beyond the scope of this paper.

This research was supported in part by NSF Grant PHY-9970771.

\listrefs

\end